\documentclass[preprint2]{aastex62}
\usepackage{natbib}
\usepackage{hyperref}
\usepackage{amsmath}
\usepackage[mathlines]{lineno}
\usepackage{multirow}
\usepackage{bm}
\usepackage{graphicx}

\begin{document}

\title{Possible discrimination of black hole origins from the lensing rate of DECIGO and B-DECIGO sources}


\author{Bin Liu$^{1,2}$, Zhengxiang Li$^{1}$, Shaoxin Zhao$^{1}$, Huan Zhou$^{3}$, and He Gao}

\affiliation{Department of Astronomy, Beijing Normal University, Beijing 100875, China; zxli918@bnu.edu.cn\\
$^2$Department of Physics and Astronomy, University of California, Irvine, CA 92697, USA\\
$^3$School of Physics and Astronomy, Sun Yat-sen University, Zhuhai, 519082, China}

\begin{abstract}
In this paper, we forecast the expected detection rates and redshift distributions of gravitationally lensed gravitational waves (GWs) from three different mass distributions of primordial black holes (PBHs) and two stellar formation models of astrophysical black holes (ABHs) in the context of DECi-hertz Interferometer Gravitational wave Observatory (DECIGO) and it's smaller scale version B-DECIGO. It suggests that DECIGO will be able to detect $10^4-10^5$ GW signals from such binary black holes (BBHs) each year and the event rate distributions for PBHs will differ from those for ABHs due to their different merger rate with respect to redshift. The large number of event rates make $5-70$ detections of lensed GW signals being possible. After considering the gravitational lensing effect, the difference between the detection rates and distributions for PBHs and ABHs will be more significant. Therefore, this can be served as a complementary method to distinguish PBHs from ABHs.    
\end{abstract}

\keywords{primordial blak holes, gravitational lensing, gravitational wave}
\section{Introduction}
Up to now, ground-based detectors have reported about ninety gravitational wave (GW) events and most of them are from the coalescence of binary black holes (BBHs). These detections are very helpful for exploring various aspects of astrophysics such as testing general relativity~\citep{Abbott2021,Abbott2019,Abbott2019b} and constraining the Hubble constant~\citep{Abbott2021b,Abbott2017}. However, the origin of these binary black holes is still an open issue. Since the first detection of GW emitted from the coalescence of BBHs by the Advanced LIGO~\citep{2016PhRvL.116f1102A}, the possibility that this signal was emitted from binary primordial black holes (PBHs) coalescence~\citep{2016PhRvL.117f1101S,2016PhRvL.116t1301B,2017PDU....18..105C,2021PhRvD.103h4001K} instead of astrophysical black holes (ABHs) was widely investigated. PBHs could be the results of the gravitational collapse of primordial fluctuations in the early universe~\citep{1971MNRAS.152...75H,1974MNRAS.168..399C},  or other mechanisms such as the quantum fluctuation~\citep{Clesse2015,Fu2019,Chen2019,Motohashi2020}, the bubble collisions~\citep{Hawking1982}, the cosmic string~\citep{Hawking1989,Hogan1984}, and the domain wall~\citep{Caldwell1996}. The diversity of formation mechanism means that PBHs might have a different evolutionary history with respect to astrophysical black holes.  

The successful detection of GW with current ground-based detectors has significantly boosted this field to the new era. Several next generation detectors, including both ground and space mission, will be put into operation in the near future with higher sensitivity. Meanwhile, covering wider frequency ranges by different facilities is also an important goal. The DECihertz Interferometer Gravitational wave Observatory (DECIGO)~\citep{Seto2001,Kawamura2019} is a future plan of a Japanese space mission for observing GWs around $f\sim 0.1\;\rm{Hz}$ and a small scale version called B-DECIGO is planned to be launched within this decade. 
DECIGO will be able to detect the GW signals of inspiralling double compact objects (DCOs) and will receive these signals earlier than ground based detectors due to the lower detection frequency range (mHz to 100 Hz). It will also has a higher sensitivity than the next generation ground based detectors. Thus, a large number of GW events are expected to be detected by this detector. It is shown that the third generation ground-based detector, i.e. the Einstein Telescope (ET), will detect $10^5-10^6$ DCO GW events per year~\citep{Piorkowska2013,Biesiada2014,Ding2015}. With a higher sensitivity, DECIGO is expected to detect the same order or even more GW events.

Given the large number of detections and further reach of the detection range, we can expect some GW events will be gravitationally lensed. LIGO-Virgo collaboration has tried to search the lensed GWs from current available observations, and concluded a null search result in the first half of O3 data~\citep{LVC2021}. Meanwhile, it was argued that, with certain  prior, the lensed GW has already been detected~\citep{Diego2021}. In any case, we will be able to accumulate numbers of lensed GW signals in the future with the help of next generation detectors. It was shown that the lensed BH-BH systems could be observed at the rate of 50 per year in DECIGO and a few in B-DECIGO~\citep{2021ApJ...908..196P}. 

In this paper, we calculated the predicted event rates of strongly lensed GW signals from PBHs and ABHs detected by DECIGO and B-DECIGO, and explored the differences in their event rates and redshift distributions. This can serve as an alternative method to distinguish PBHs and ABHs. This paper is organized as follows. In section~\ref{sec: Merger rate density of PBHs and ABHs}, we introduce three theoretical mass distributions of PBH and two different stellar formation models of ABH used in this paper. In section~\ref{sec: Detection rate for unlensed events}, we calculate the detection rate of unlensed events. 
We discuss lensed events in section \ref{sec: Detection rate for lensed events}. Finally, we summarize our results and draw conclusions in section \ref{sec: Conclusion}. Throughout this paper, we adopt a flat ${\Lambda}$CDM cosmological model with $H_0 = 70$ km s$^{-1}$ Mpc$^{-1}$ and $\Omega_m = 0.3$.   

\section{Merger rate density of PBHs and ABHs}\label{sec: Merger rate density of PBHs and ABHs}

To calculate the observable events for ABH models, we take the merger rate of a convolution of the birthrate of ABHs as
\begin{equation}\label{eq2-1}
\begin{split}
\mathcal{R}_{\rm ABH}(z,m_1)=\int_{t_{\rm min}}^{t_{\rm max}}R_{\rm b}(t(z)-t_{\rm d}, m_1)\\
\times P(t_{\rm d})dt_{\rm d},
\end{split}
\end{equation}
where the minimum time delay is set as $t_{\rm min}$ $=50~\rm Myr$, $t_{\rm max}$ is set as the Hubble time, $t_{\rm d}$ is the time delay, and $t(z)$ is the age of the universe at the moment of merger. In addition, $R_{\rm b}(t, m_1)$ can be estimated from~\citep{Dvorkin:2016wac}
\begin{equation}\label{eq2-2}
\begin{split}
R_{\rm b}(t, m_{\rm bh})=\int_0^{\infty}\psi[t-\tau(m')]\phi(m')\\
\times \delta(m'-g^{-1}_{\rm b}(m_{\rm bh}))dm',
\end{split}
\end{equation}
where $m_{\rm bh}$ is the mass of the remnant black hole, $\tau(m)$ is the lifetime of a progenitor star, and $\phi(m)\propto m^{-2.35}$ is the initial mass function. In order to get $g^{-1}_{\rm b}(m_{\rm bh})$, we should consider the relation as
\begin{equation}\label{eq2-3}
\begin{split}
\frac{m_{\rm bh}}{m}=A\bigg(\frac{m}{40~M_{\odot}}\bigg)^{\beta}\frac{1}{\bigg(\frac{Z}{0.01~Z_{\odot}}\bigg)^{\gamma}+1},
\end{split}
\end{equation}
where $Z$ is the metallicity~\citet{Belczynski:2016obo}. The values of parameters $[A,\beta,\gamma]$ are taken as $[0.3,0.8,0.2]$~\citep{Dvorkin:2016wac}. The star formation rate $\psi(t)$ in $R_{\rm b}(t, m_1)$ is given by
\begin{equation}\label{eq2-4}
\psi(z)=k\frac{a~\exp[b(z-z_m)]}{a-b+b~\exp[a(z-z_m)]},
\end{equation}
where the parameters $[k,a,b,z_m]$ are dependent on the specific ABH formation scenarios. Here, we consider two ABH models. The first one is the Fiducial model, which corresponds to the classical isolated binary evolution of ABH. The parameters $\{k,a,b,z_m\}$ for the Fiducial model are from~\citet{Vangioni:2014axa}, i.e. $[0.178~M_{\odot}\rm{yr}^{-1}\rm{Mpc}^{-3}, 2.37, 1.80, 2.00]$. The second one is the GRB-based model, which uses the SFR calibrated from the GRB at high redshift. The values for these parameters are set as  $[0.146~M_{\odot}\rm{yr}^{-1}\rm{Mpc}^{-3}, 2.80, 2.46, 1.72]$~\citep{Vangioni:2014axa}. For local merger rate $R_{0,\rm ABH}\equiv\int_{\geq1~M_{\odot}}\mathcal{R}_{\rm ABH}(z,m_1)dm_1$, we take it as $23.9^{+14.3}_{-8.6}~\rm Gpc^{-3}yr^{-1}$ in~\citet{Abbott2021c}.

In order to obtain the GW event rates of PBH systems, their intrinsic merger rate must be calculated first as well. 
For a general mass function $P(m)$, the time dependent comoving merger rate density is~\citep{Chen2018}:
\begin{equation}\label{eq1}
\resizebox{0.95\hsize}{!}{$\begin{split}
\mathcal{R}_{\rm PBH}(m_1,m_2,z|\theta,f_{\rm PBH})=2.8\times10^6\bigg(\frac{t(z)}{t_0}\bigg)^{-\frac{34}{37}}f_{\rm PBH}^2\\
(0.7f_{\rm PBH}^2+\sigma_{\rm eq}^2)^{-\frac{21}{74}}\times
{\rm min}\bigg(\frac{P(m_1|\theta)}{m_1},\frac{P(m_2|\theta)}{m_2}\bigg)\\
(m_1m_2)^{\frac{3}{37}}(m_1+m_2)^{\frac{36}{37}}\bigg(\frac{P(m_1|\theta)}{m_1}+\frac{P(m_2|\theta)}{m_2}\bigg),
\end{split}$}
\end{equation}
in units of $\rm M_{\odot}^{-2}Gpc^{-3}yr^{-1}$, where $\theta$ represents the parameter vector for each mass distribution model. $f_{\rm PBH}=\frac{\Omega_{\rm PBH}}{\Omega_{\rm CDM}}$ is the abundance of PHBs in the dark matter at present universe. $\sigma_{ep}$ is the variance of the rest dark matter density perturbations at scale $\mathcal{O}(10^0-10^3)M_{\odot}$ and we usually take the value of it as $\sigma_{eq} \approx 0.005$~\citep{Chen2018,Hamoud2017}. By integrating over the mass, we can get the merger rate of all PBHs at cosmic time $t$:
\begin{equation}
\resizebox{0.95\hsize}{!}{$\begin{split}
R_{\rm PBH}(z|\theta,f_{\rm PBH})=\int_{0}^{\infty}\int_{0}^{m_1} \mathcal{R}_{\rm PBH}(m_1,m_2,z|\theta,f_{\rm PBH})
d m_{2} d m_{1}
\end{split}$}
\end{equation}
and the local merger rate is $R_{0,\rm PBH} \equiv R_{\rm PBH}(z=0|\theta,f_{\rm PBH})$. Here we adopt three mass functions of PBHs. These are the power law mass function, the log-normal mass function, and the critical collapse mass function.
The power law mass function can be written as:
\begin{equation}
P\left(m, \alpha, M_{\min }\right)=\frac{\alpha-1}{M_{\min }}\left(\frac{m}{M_{\min }}\right)^{-\alpha}
\end{equation}
with $m\geq M_{min}$ and $\alpha>1$. This mass function is a natural prediction when PBHs are generated by scale-invariant density fluctuations or from the collapse of cosmic strings. The log-normal mass function is expressed as~\citep{Hutsi2021,Bellomo2018,Carr2017}:
\begin{equation}
P\left(m, \sigma, m_{\mathrm{c}}\right)=\frac{1}{\sqrt{2 \pi} \sigma m} \exp \left(-\frac{[\ln \left(m / m_{\mathrm{c}}\right)]^2}{2 \sigma^{2}}\right)
\end{equation}
where $m_c$ and $\sigma$ denote the peak mass of $mP(m)$ and the width of mass spectrum, respectively. This mass function is often a good approximation if the PBHs are produced from a smooth symmetric peak in the inflationary power spectrum. And the last mass function - the critical collapse distribution is~\citep{Hutsi2021,Carr2016,Carr2017}:
\begin{equation}
P\left(m, M_{f}\right)=\frac{\gamma^{2}}{M_{f}^{1+\gamma} \Gamma(1 / \gamma)} m^{\gamma} \exp \left(-\left(\frac{m}{M_{f}}\right)^{\gamma}\right)
\end{equation}
where $M_f$ is a mass-scale that corresponds to the horizon mass at the collapse epoch, $\gamma$ is a universal exponent which is related to the critical collapse of radiation, and~$\Gamma(x)$ denotes the gamma function. This mass function is supposed to be closely related to PBHs originating from density fluctuations with a $\delta$-function power spectrum.

To extract the population parameters $\{\theta, f_{\rm PBH}\}$ for PBH models from $N_{\rm obs}$ detections of GW events $d = \{d_1,...d_{N_{\rm obs}}\}$, we can perform the hierarchical Bayesian inference~\citep{Mandel:2018mve,Chen:2018rzo,Chen:2019irf,Wu:2020drm,Hutsi2021}. Firstly, the likelihood for $N_{\rm obs}$ BBH events is  
\begin{equation}\label{eq2-5}
\begin{split}
p(d|\theta,f_{\rm PBH})=(N(\theta,f_{\rm PBH}))^{N_{\rm obs}}e^{-N(\theta,f_{\rm PBH})\xi(\theta,f_{\rm PBH})}\\
\times \prod_{i}^{N_{\rm obs}}\int d\lambda L(d_i|\lambda)p_{\rm pop}(\lambda|\theta,f_{\rm PBH}),
\end{split}
\end{equation}
where $\lambda\equiv\{m_1,m_2,z\}$, the likelihood of one BBH event $L(d_i|\lambda)$ is proportional to the posterior $p(\lambda|d_i)$, and $N(\theta,f_{\rm PBH})$ is the total number of events in the model characterised by the set of population parameters $\{\theta, f_{\rm PBH}\}$ as
\begin{equation}\label{eq2-6}
\begin{split}
N(\theta,f_{\rm PBH})=\int d\lambda T_{\rm obs}\mathcal{R}_{\rm PBH}(\lambda|\theta, f_{\rm PBH})\frac{1}{1+z}\frac{dV_{\rm c}}{dz}
\end{split}
\end{equation}
where $dV_{\rm c}/{dz}$ is the differential comoving volume, the factor $1/(1 + z)$ accounts for the cosmological time dilation from source frame to the detector frame, and $T_{\rm obs}$ is effective observing time of LIGO O1-O3. $p_{\rm pop}(\lambda|\theta,f_{\rm PBH})$ is the distribution of BH masses and redshifts in coalescing binaries as
\begin{equation}\label{eq2-7}
\begin{split}
p_{\rm pop}(\lambda|\theta,f_{\rm PBH})=\frac{1}{N(\theta,f_{\rm PBH})}\times\\
\bigg[T_{\rm obs}\mathcal{R}_{\rm PBH}(\lambda|\theta, f_{\rm PBH})
\frac{1}{1+z}\frac{dV_{\rm c}}{dz}\bigg],
\end{split}
\end{equation}
Meanwhile, $\xi(\theta,f_{\rm PBH})$ can be calculated by 
\begin{equation}\label{eq2-8}
\xi(\theta,f_{\rm PBH})\approx\frac{1}{N_{\rm inj}}\sum_{k=1}^{N_{\rm det}}\frac{p_{\rm pop}(\lambda_k|\theta,f_{\rm PBH})}{p_{\rm draw}(\lambda_k)},
\end{equation}
where $N_{\rm inj}$ is the total number of injections, $N_{\rm det}$ is the number of injections that are successfully detected, and $p_{\rm draw}$ is the probability distribution from which the injections are drawn. Then the posterior probability distribution $p(\theta,f_{\rm PBH}|d)$ can be calculated from
\begin{equation}\label{eq2-9}
p(\theta,f_{\rm PBH}|d)=p(d|\theta,f_{\rm PBH})p(\theta,f_{\rm PBH}),
\end{equation}
where $p(\theta,f_{\rm PBH})$ is the prior distribution for $\theta$ and $f_{\rm PBH}$. We use the flat and log-uniform prior for the $\theta$ and $f_{\rm PBH}$, respectively. 

We adopt the available BBHs data from GWTC-3~\citep{LIGOScientific:2021djp}, these GW data satisfy the following criteria: black hole masses ($m_1$ and $m_2$)  larger than $3~M_{\odot}$, effective spin ($|\chi_{\rm eff}|$) lower than $0.3$, network-matched filter signal-to-noise ratio ($\rm SNR$) larger than 9 and inverse false alarm rate ($\rm ifar$) higher than 1 year. Totally, there are 58 events from LIGO's O1, O2,  O3a and O3b runs. Based on these data, we incorporate the PBH population distribution into the \texttt{ICAROGW} to estimate the likelihood function and use \texttt{Bilby} to search over the parameter space~\citep{Mastrogiovanni:2021wsd,Ashton:2018jfp}. Finally, the results of population parameters $\{\theta, f_{\rm PBH}\}$ of different PBH models are shown in Table~\ref{tab:PBH mass parameter and meger rate}. Although, there are still many discussions about the origin mechanism of GWs of BBH. If the range of fraction of PBH in DM at present universe can be from $10^{-3}$ to $10^{-2}$, the merging model of PBH binaries (Equation~(\ref{eq1})) can provide enough BBH events to explain the case theoretically~\citep{Chen2018}. Furthermore, possibilities of the above merging model of PBH binaries have been explored by combining the PBH population and GW events detected by LIGO-Virgo  \citep{Chen:2018rzo,Chen:2019irf,Wu:2020drm}. In addition, some studies suggest that the current GW data prefers the binaries model of mixed PBH and ABH populations~\citep{Hutsi2021,DeLuca:2021wjr,Franciolini:2021tla}. Therefore, this may require more GW data to discriminate in the future.

\begin{table*}[htbp]
  \center
  \caption{The resluts for population parameters $\{\theta, f_{\rm PBH}\}$ of different PBH models and best fit value of local merger rate of three PBH models} \label{tab:PBH mass parameter and meger rate}
    \begin{tabular}{ccccc}
    \hline
    Mass function & Parameter~1 & Parameter~2 ($M_{\odot}$) & $\log_{10}f_{\rm PBH}$ & $R_{0, \rm PBH}~(\rm Gpc^{-3}yr^{-1})$  \\
   \hline
    Power law    & $\alpha = 1.91^{+0.10}_{-0.11}$ & $M_{min} = 6.43^{+0.30}_{-0.45}$ & $-2.74^{+0.03}_{-0.03}$ & 22\\
    Log-normal & $\sigma=0.76^{+0.08}_{-0.06}$   &$m_c=16.78^{+1.57}_{-1.54}$       & $-2.81^{+0.04}_{-0.04}$ & 20 \\
    Critical collapse & $\gamma=0.86^{+0.12}_{-0.12}$        &  $M_f=8.13^{+2.32}_{-2.32}$     & $-2.80^{+0.04}_{-0.04}$& 24\\
    \hline
    \end{tabular}
\end{table*}

\begin{figure}[htbp]
     \centering
     \includegraphics[width=0.48\textwidth]{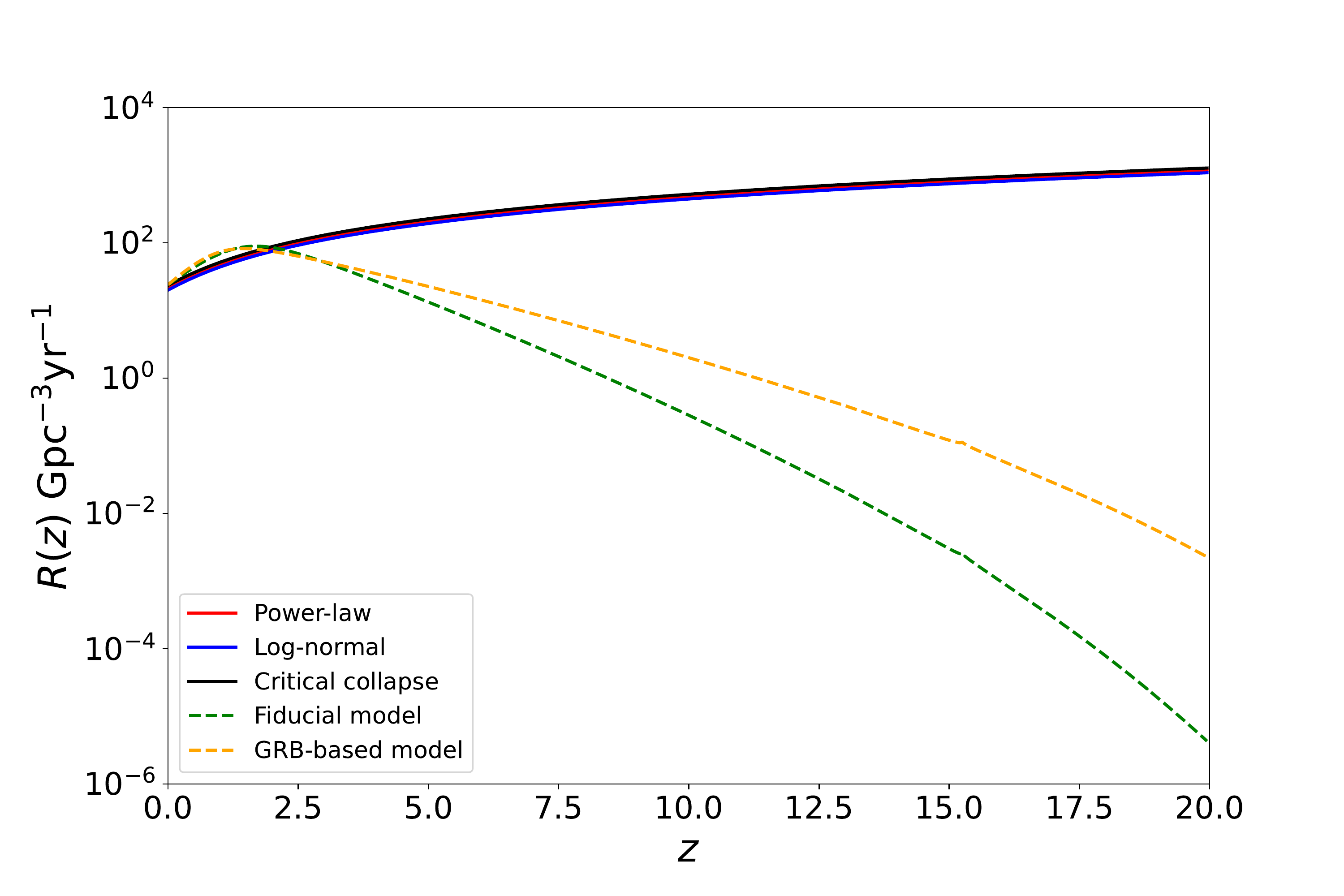}
     \caption{Merger rate $R(z)$ of different BBH scenarios as a function of redshift. The solid lines illustrate the results of three PBH models and the dash lines show the results of two ABH models. }
     \label{DECIGO merger rate}
\end{figure}

\section{Detection rate for unlensed events}\label{sec: Detection rate for unlensed events}

As shown in Figure~\ref{DECIGO merger rate}, the merger rates of binary PBHs and ABHs evolve differently with respect to redshift, especially at high redshift. By studying the redshift distribution of GW events, we can distinguish PBHs and ABHs. We first calculated the unlensed event rates of these sources. The yearly detection rate of the DCOs at redshift $z_s$ with signal-to-noise (SNR) exceeding the detector's threshold is: 
\begin{equation}
\label{eq: event rate}
\dot{N} (>\rho_0 | z_s)=\int_{0}^{z_s} \frac{d\dot{N} (>\rho_0)}{dz} dz
\end{equation}
where
\begin{equation}
\begin{split}
\frac{d\dot{N} (>\rho_0)}{dz_s} = 4\pi \bigg(\frac{c}{H_0}\bigg)^3 \frac{{R}_0 (z_s)}{1+z_s} \frac{\tilde{r}^2 (z_s)}{E(z_s)} \\
\times C_{\Theta}(x(z_s,\rho_0))
\end{split}
\end{equation}
Here we set the threshold $\rho_0=8$. ${R}_0 (z_s)$ denotes the intrinsic merger rate of ABHs or PBHs at redshift $z_s$ as mentioned above. $\tilde{r}^2 (z_s) = \int^{z}_{0} \frac{z^{\prime}}{E(z^{\prime})}$ is the non-dimensional comoving distance with $E(z) = H(z)/H(0)$. $C_{\Theta}(x)=\int_{x}^{\infty} P_{\Theta} d\Theta$ with the probability distribution of $\Theta$, $P_{\Theta}(\Theta)=5\Theta(4-\Theta)^3/256$ for $0<\Theta<4$. $x(z,\rho)=\frac{\rho}{8} (1+z)^{1/6} \frac{c}{H_0}\frac{\tilde{r}(z)}{r_0}\big(\frac{1.2M_{\odot}}{\mathcal{M}_0}\big)^{5/6}$ which is correlated with the detector's characteristic distance parameter $r_0$. This parameter depends on the noise spectrum of the detector and can be estimated as 
\begin{equation}
    \label{eq:r0}
    r_{0}^{2} = \frac{5}{192 \pi^{4/3}} \bigg(\frac{3}{20}\bigg)^{5/3} \frac{(GM_{\odot})^{5/3}}{c^3} f_{7/3}.
\end{equation}
where
$f_{7/3}=\int_{0}^{\infty} [f^{7/3} S_h (f)]^{-1} df$.
The noise spectrum $S_h(f)$ of DECIGO is \citep{2011CQGra..28i4011K}:
\begin{equation}
\begin{split}
    S_h(f) = 10^{-48}\times \bigg[ 7.05\bigg(1+\frac{f^2}{f_{p}^{2}}\bigg)+4.8 \times 10^{-3} \\
    \times \frac{f^{-4}}{1+\frac{f^2}{f_{p}^{2}}}+5.33\times10^{-4}f^{-4}\bigg] \rm{Hz^{-1}}
\end{split}
\end{equation}
where $f_p=7.36\;\rm{Hz}$. And the noise spectrum of B-DECIGO is
\begin{equation}
    \begin{split}
        S_h(f)=10^{-46}\times \bigg[4.040+6.399\times10^{-2}f^{-4}\\ +6.399\times10^{-3}f^2\bigg]\rm{Hz^{-1}}
    \end{split}
\end{equation}
The confusion noise from unresolved DCO systems will influence the detection results, so we include the noise spectrum caused by these systems as well. The sky-averaged spectrum of these GW backgrounds can be expressed as:

\begin{equation}
S_h^{\mathrm{GW}, \mathrm{DCO}}=\frac{4}{\pi} f^{-3} \rho_{\mathrm{cr}} \Omega_{\mathrm{GW}}^{\mathrm{DCO}}
\end{equation}

where $\rho_{cr}$ is the critical density of the universe and $\Omega^{\rm DCO}_{\rm GW}$ is the energy density parameter corresponding to the unresolved signals from the DCO systems. (refer to~\citealt{2021ApJ...908..196P} for details). The total spectrum is $S_h(f)+S_h^{\rm{GW,DCO}}(f)$.  Substituting the total spectrum into equation~(\ref{eq:r0}), the final characteristic distance $r_0$ is different in different BBH scenarios. Table~\ref{table:$r_0$} listed the results of DECIGO and B-DECIGO with the BBH systems used in this paper.

\begin{table}[h!]
    \caption{Characteristic distance parameters of DECIGO ($r_0$) and B-DECIGO ($r_{\rm 0b}$), including the influences of unresolved background spectrum in each BBH scenario. }
    \label{table:$r_0$}
    \begin{center}
    \begin{tabular}{lcc}
    \hline
    BBH scenario            & $r_0$ (Mpc)    & $r_{\rm 0b}$ (Mpc) \\
    \hline
    fiducial model          & 990                   & 404  \\
    GRB based model         & 981                   & 402  \\
    power law model         & 899                   & 382  \\
    log normal model        & 928                   & 389  \\
    critical collapse model & 879                   & 377  \\
    \hline
    \end{tabular}
    \end{center}
\end{table}

Combining the detector characteristic distance and intrinsic merger rate for a given model, we can calculate the redshift distributions of observed BBH events for both DECIGO and B-DECIGO. The results are shown in Figure~\ref{DECIGO events} and Figure~\ref{B-DECIGO events}. As shown in Figure~\ref{DECIGO events}
ABHs and PBHs follow different redshift distributions. That is, the ABH events assemble at lower redshifts while the PBH events have a more flattened shape and have more detectable events at higher redshifts. The total yearly event rates were also estimated from Equation~(\ref{eq: event rate}). The results are shown in Table~\ref{no-lensed detection rate}. It suggests that there are significant differences between the event rates of ABH models and the ones of PBH models. We will expect to observe an order of magnitude more PBH events than ABH. For B-DECIGO, due to the smaller characteristic distance, fewer high redshift merger events could be detected than DECIGO. Though the differences are not as significant as DECIGO, we can still distinguish the distributions of ABHs and PBHs in the redshift range $z<8$ as shown in Figure~\ref{B-DECIGO events}.

\section{Detection rate for lensed events}\label{sec: Detection rate for lensed events}
With numerous DCO GW events supposed to be detected, we naturally expect some of such events to be gravitationally lensed by intervening galaxies. The lensing optical depth can be expressed as:

\begin{equation}
\tau_{\pm}= \frac{16}{30} \pi^3 \bigg(\frac{c}{H_0}\bigg)^3 \tilde{r}(z_s)^{3} \bigg(\frac{\sigma_{\ast}}{c}\bigg)^4 n_{\ast} \frac{\Gamma (\frac{4+\alpha}{\beta})}{\Gamma (\frac{\alpha}{\beta})} y_{\pm,max}^2.
\end{equation}

\begin{figure}[htbp]
     \centering
     \includegraphics[width=0.48\textwidth]{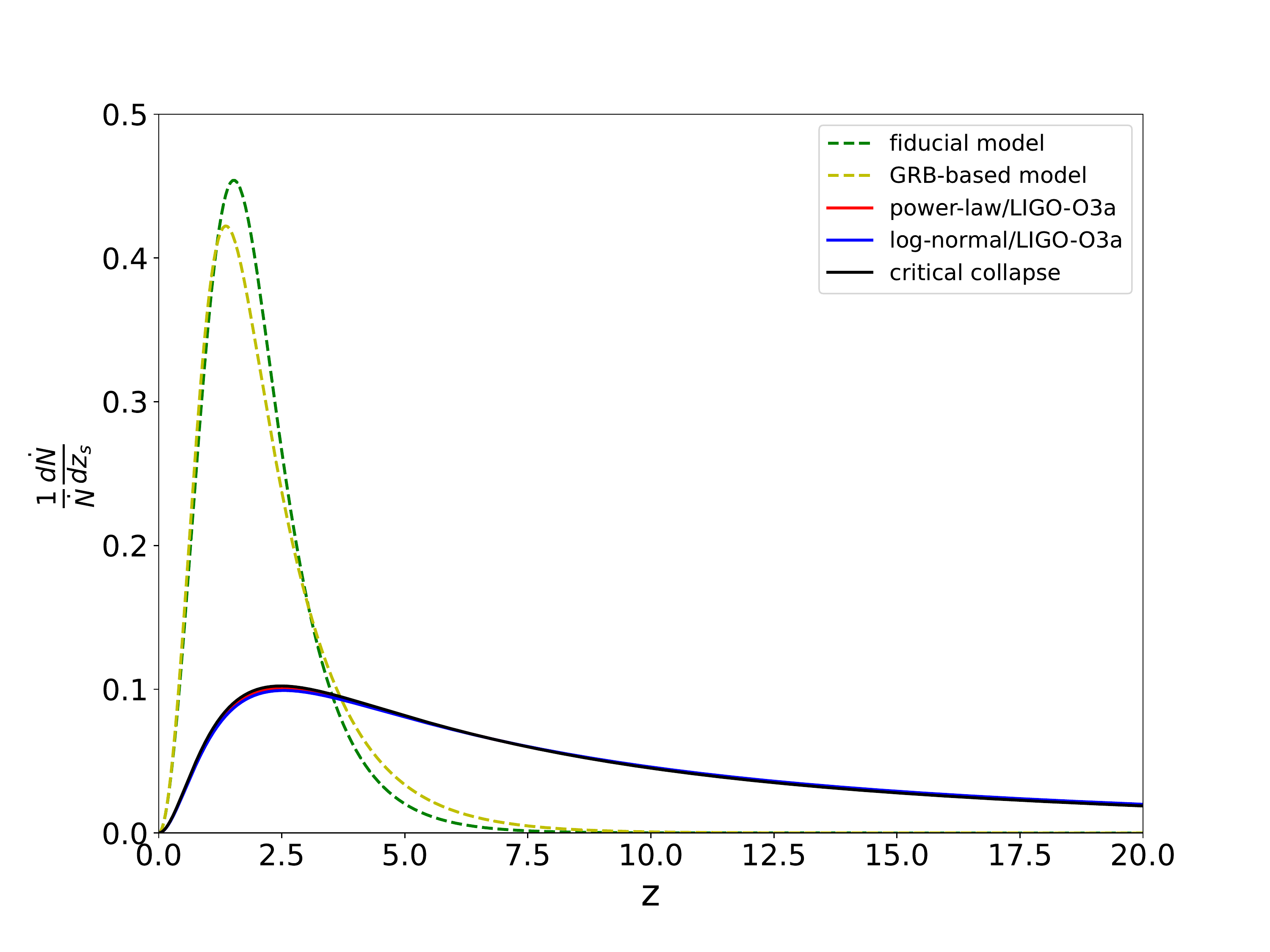}
     \caption{Probability density distribution of BBH events to be detected by DECIGO as a function of redshift.}
     \label{DECIGO events}
\end{figure}

\begin{figure}[htbp]
     \centering
     \includegraphics[width=0.48\textwidth]{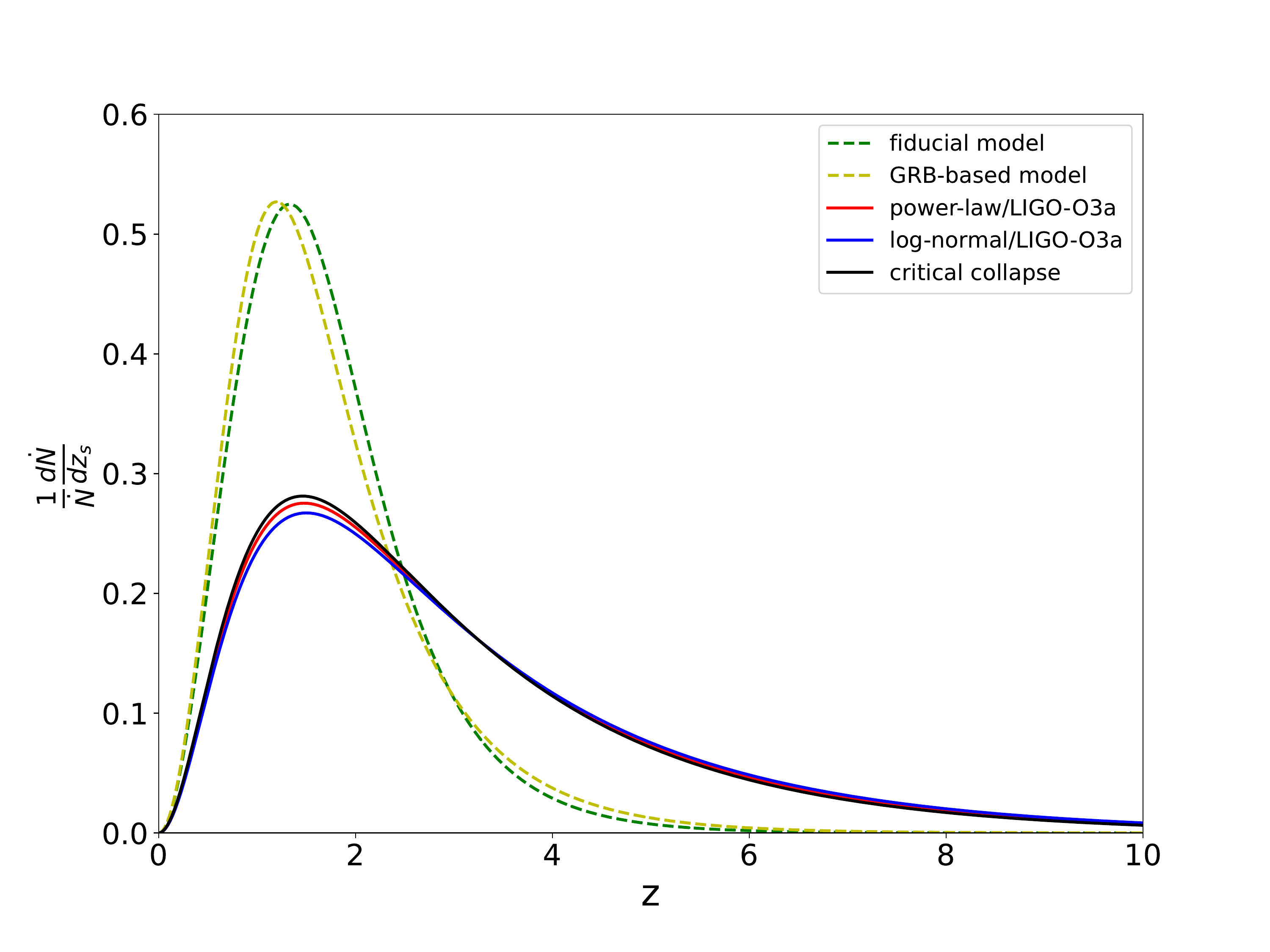}
     \caption{Probability density distribution of BBH events to be detected by B-DECIGO as a function of redshift.}
     \label{B-DECIGO events}
\end{figure}

\begin{table}[ht!]
    
    \caption{Predicted yearly detection rate of different BBH models for DECIGO and B-DECIGO}
    \label{no-lensed detection rate}
    \centering
    \begin{tabular}{lcc}
    \hline
    Yearly detection rate       &  DECIGO     &  B-DECIGO    \\
    \hline
    fiducial model              &  31146.5      &  18854.7   \\
    GRB based model             &  31771.0      &  18700.8   \\
    power law model             &  115889.6     &  24552.9   \\
    log normal model            &  108541.5     &  23401.2   \\
    critical collapse model     &  121146.7     &  25358.7   \\
    \hline
    \end{tabular}
    
\end{table}

We assume conservatively that the lensing galaxies are all elliptical galaxies and can be characterized by the singular isothermal spheres (SIS) model.
The parameters of velocity dispersion distribution function are then taken as $\sigma_{*}=161\pm5\ {\rm km/s}$, $n_{*}=8.0\times 10^{-3}\ {\rm h}^3\ {\rm Mpc}^{-3}$, $\alpha=2.32\pm0.10$, $\beta=2.67\pm0.07$ \citep{Choi2007}. $\tau_{+}$ and $\tau_{-}$ denote the optical depth of brighter image and fainter image, respectively. $y$ is the misalignment of the source with respect to the optical axis of the lens and is related to the magnification factors by $\mu_{\pm}=\frac{1}{y}\pm 1$. Assuming the detection threshold $\rho_0=8$, the misalignment should not exceed the limiting value  $y_{\pm,max}=[({8\over \rho_{intr}})^2\mp 1]^{-1}$  (see ~\citealt{Piorkowska2013,Biesiada2014,Ding2015} for details).

\begin{figure}[htbp]
     \centering
     \includegraphics[width=0.48\textwidth]{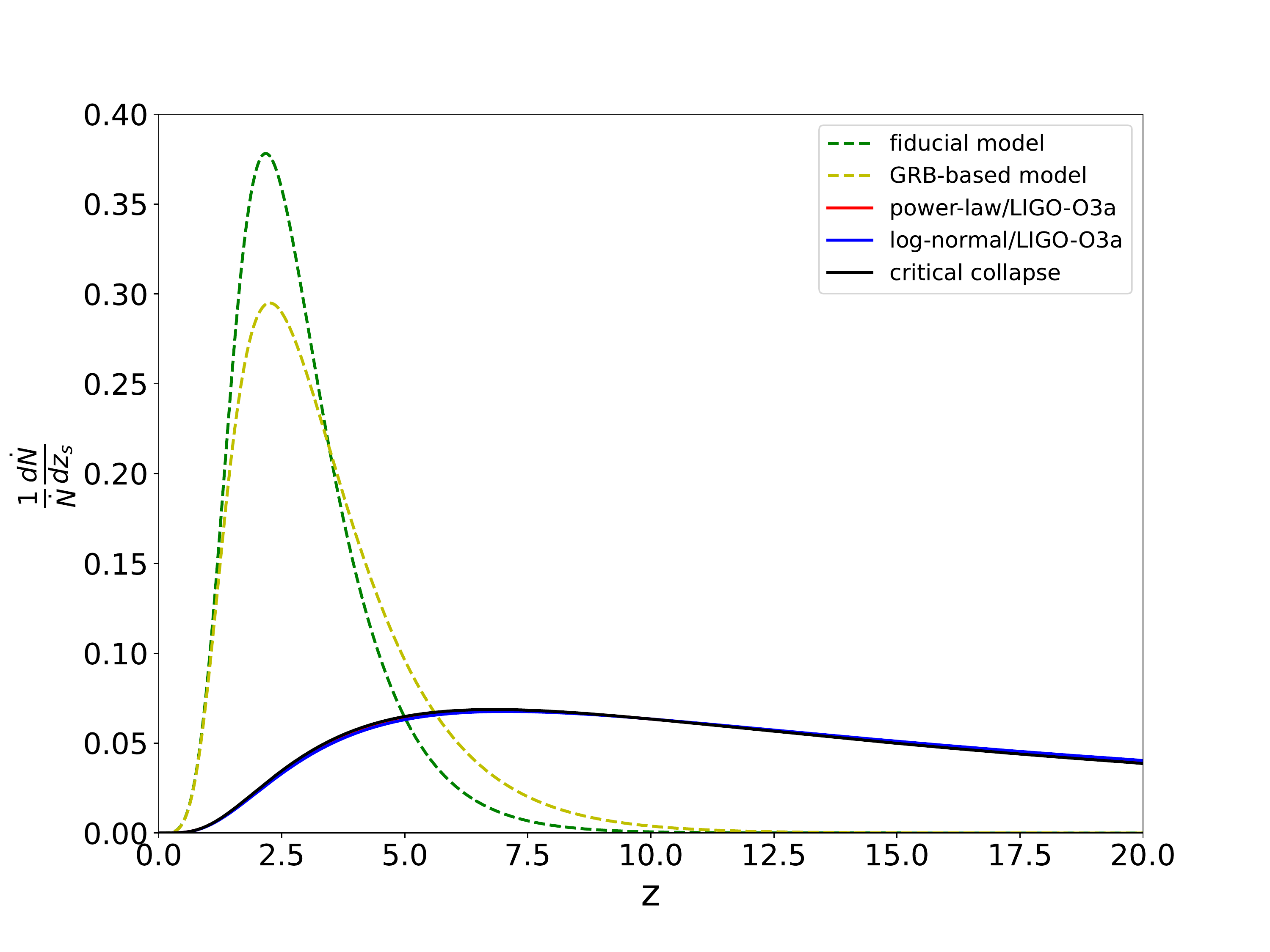}
     \caption{Probability density distribution of lensed BBH events to be detected by DECIGO as a function of redshift.}
     \label{Lensde DECIGO events}
\end{figure}

\begin{figure}[htbp]
     \centering
     \includegraphics[width=0.48\textwidth]{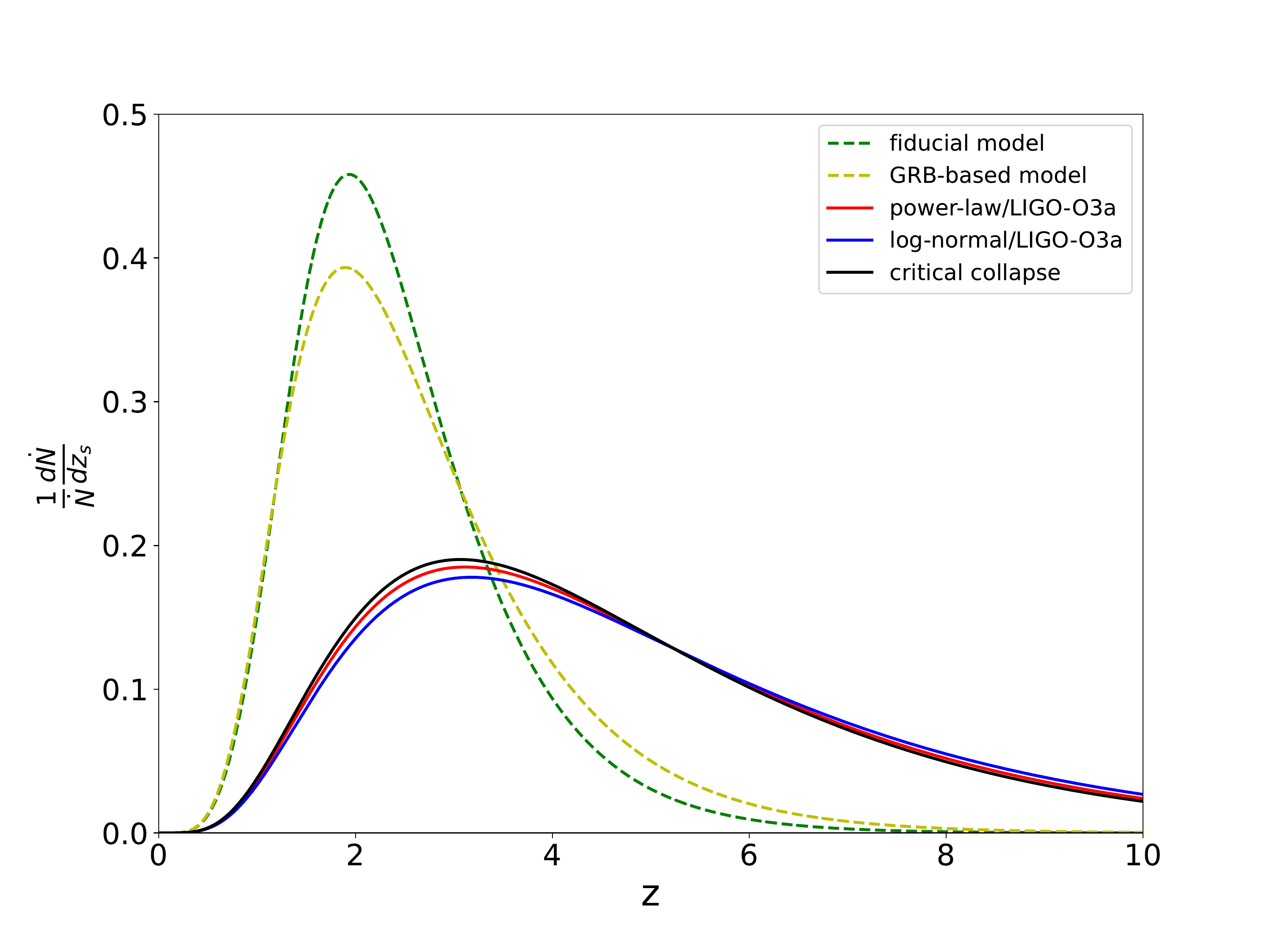}
     \caption{Probability density distribution of lensed BBH events to be detected by B-DECIGO as a function of redshift.}
     \label{Lensed B-DECIGO events}
\end{figure}

The intrinsic low SNR ($\rho_{int} < 8$) signal will be detectable due to the lensing magnification effect, this will increase the detection rate to: 
\begin{equation}
\frac{\partial{^2 \dot N}}{\partial{z_s}\partial{\rho}} = 4\pi \bigg(\frac{c}{H_0}\bigg)^3 \frac{R_0 (z_s)}{1+z_s} \frac{\tilde{r}^2 (z_s)}{E(z_s)} P_{\Theta}(x(z_s,\rho))
\frac{x(z_s,\rho)}{\rho}
\end{equation}
Combing this detection rate with the lensing depth, the rate of lensed GW events are:
\begin{equation}
    \dot{N}_{lensed} = \int^{z_s}_{0}dz_s \int^{\rho_0}_{0}\tau(z_s,\rho)\frac{\partial{^2 \dot N}}{\partial{z_s}\partial{\rho}}d\rho
\end{equation}
The total detection rate includes both the case with weaker intrinsic signal $\rho_{int}<8$ (but exceeds 8 after lensing magnification) and the one with stronger signal $\rho_{int}>8$. 
The above calculation considers the detection as a continuous survey. In practice, detectors have a finite mission duration, and we will miss some lensing signals due to the time delay. Therefore, with this effect taken into account, the optical depth should be corrected as:

\begin{equation}
\tau_{\Delta, \pm}=\tau \bigg[1-\frac{1}{7} \frac{\Gamma (\frac{\alpha +8}{\beta})}{\Gamma (\frac{\alpha +4}{\beta})} \frac{\Delta t_{\ast,\pm}}{T_{surv}}\bigg]
\end{equation}
where $\Delta t_{*,\pm} = \frac{32\pi^2}{H_0}\tilde{r}(z_s)(\frac{\sigma_*}{c})^4y_{\pm,max}$ and $T_{surv}$ is the mission duration of the telescope. 
Here we obtain the event rates of the lensed GW of PBHs and ABHs by considering $T_{surv}=4$ years as planned by DECIGO and B-DECIGO. The results are shown in Table~\ref{lensed detection rate}. The differential lensing rates are also illustrated in Figure~\ref{Lensde DECIGO events} and Figure~\ref{Lensed B-DECIGO events}. The high redshift proportion of PBHs detected by both DECIGO and B-DECIGO will increase as shown in Figure \ref{DECIGO events}, \ref{Lensde DECIGO events} and Figure \ref{B-DECIGO events}, \ref{Lensed B-DECIGO events}. So the distributions of ABH and PBH events are more distinguishable. These differences are reflected in the detection numbers as well. The lensed signals have a higher ratio between PBH and ABH than no-lensed signals as shown in Table~\ref{lensed detection rate} and Table~\ref{no-lensed detection rate}. 

\begin{table}[ht!]
    
    \caption{Predicted yearly detection rate of lensed events for different BBH models with DECIGO and B-DECIGO}
    \label{lensed detection rate}
    \centering
    \begin{tabular}{lcc}
    \hline
    Yearly detection rate       &  DECIGO                 &  B-DECIGO  \\
    \hline
    fiducial model              &  4.6                    &  2.2   \\
    GRB based model             &  5.2                    &  2.2   \\
    power law model             &  68.8                  &  6.1   \\
    log normal model            &  65.0                   &  6.1   \\
    critical collapse model     &  71.4                   &  6.2   \\
    \hline
    \end{tabular}
    
\end{table}

\section{Conclusion}\label{sec: Conclusion} 
In this paper, we yield predictions for the detection rates of the GW signals from coalescences of BBHs using DECIGO and B-DECIGO. These will provide us a complementary tool to distinguish the ABHs and PBHs, especially using the lensed signals.

The detection capability of DECIGO is significantly improved , and it is likely to detect $10^4-10^5$ BBHs every year. This will enable us to distinguish the differences in event rates and redshift distributions of GW from ABHs and PBHs. It suggests that the predicted event number of GW signals from the coalescence of PBHs is $\sim 4$ times larger than the one from the coalescence of ABHs. This difference implies that we can probably explore the origins of BBHs i.e. PBHs or ABHs, with the detection of GW signals from the coalescence of them. Though the redshift distributions of both ABHs and PBHs peak at $z\sim2$, the distribution of ABHs declines rapidly and ends at $z\sim6$. For the distribution of PBHs, it shows a more flattened shape and can be reached to $z>10$.

With these numerous and high redshift detections, we further calculated the results of gravitationally lensed signals detected by DECIGO. The differences between ABH and PBH scenarios are more visible. DECIGO will report $\sim 5$ and $\sim 70$ lensed GW signals from ABHs and PBHs, respectively. That is, the event rates of lensed GW signals from PBHs might become about $14$ times larger than that of ABHs. Meanwhile, the redshift distributions of lensed GW signals from PBHs will become more flattened because of the magnification effect. Moreover, the difference between ABH and PBH scenarios in high redshift range will thus be more significant.

B-DECIGO will also be able to provide differences both in the event rates and the distributions, though the differences are not as significant as DECIGO due to its lower sensitivity. The results also become more distinguishable in the lensed events. With the help of the gravitational lensing effect, we expect to explore more distant coalescences of BBHs by (B-)DECIGO and other future detectors.

\section{Acknowledgements}
We would like to thank Shaoqi Hou and Marek Biesiada for their helpful discussions. This work was supported by the National Key Research and Development Program of China Grant No. 2021YFC2203001, the National Natural Science Foundation of China under Grants Nos. 12275021, 11920101003, 11722324, 11603003, 11633001, and U1831122, the science research grants from the China Manned Space Project with No. CMS-CSST-2021-B11, the Strategic Priority Research Program of the Chinese Academy of Sciences, Grant No. XDB23040100, and the Interdiscipline Research Funds of Beijing Normal University.

\bibliography{refs}

\end{document}